\documentclass[runningheads]{llncs}
\usepackage{graphicx}
\usepackage{multirow}
\usepackage{parskip}
\begin{document}
\title{Enhancing Cardiac MRI Segmentation via Classifier-Guided Two-Stage Network and All-Slice Information Fusion Transformer}
%
%
\author{Zihao Chen\inst{1,2,3}\thanks{Contribution from Zihao Chen was carried out during his internship at United Imaging Intelligence, Cambridge, MA.} \and Xiao Chen\inst{1} \and
Yikang Liu\inst{1} \and Eric Z. Chen\inst{1} \and Terrence Chen\inst{1}
\and Shanhui Sun\inst{1}\thanks{Corresponding author: shanhui.sun@uii-ai.com}}

\institute{United Imaging Intelligence, Cambridge, MA, USA \and
University of California, Los Angeles, Los Angeles, CA, USA \and Cedars-Sinai Medical Center, Los Angeles, CA, USA}

\authorrunning{Z. Chen et al.}
\maketitle              
\begin{abstract}
Cardiac Magnetic Resonance imaging (CMR) is the gold standard for assessing cardiac function. Segmenting the left ventricle (LV), right ventricle (RV), and LV myocardium (MYO) in CMR images is crucial but time-consuming. Deep learning-based segmentation methods have emerged as effective tools for automating this process. However, CMR images present additional challenges due to irregular and varying heart shapes, particularly in basal and apical slices. In this study, we propose a classifier-guided two-stage network with an all-slice fusion transformer to enhance CMR segmentation accuracy, particularly in basal and apical slices. Our method was evaluated on extensive clinical datasets and demonstrated better performance in terms of Dice score compared to previous CNN-based and transformer-based models. Moreover, our method produces visually appealing segmentation shapes resembling human annotations and avoids common issues like holes or fragments in other models' segmentations.
\keywords{CMR segmentation  \and Classifier-guided \and Transformer.}
\end{abstract}
\section{Introduction}

\begin{figure}[t]
\centering
\includegraphics[width=0.9\textwidth]{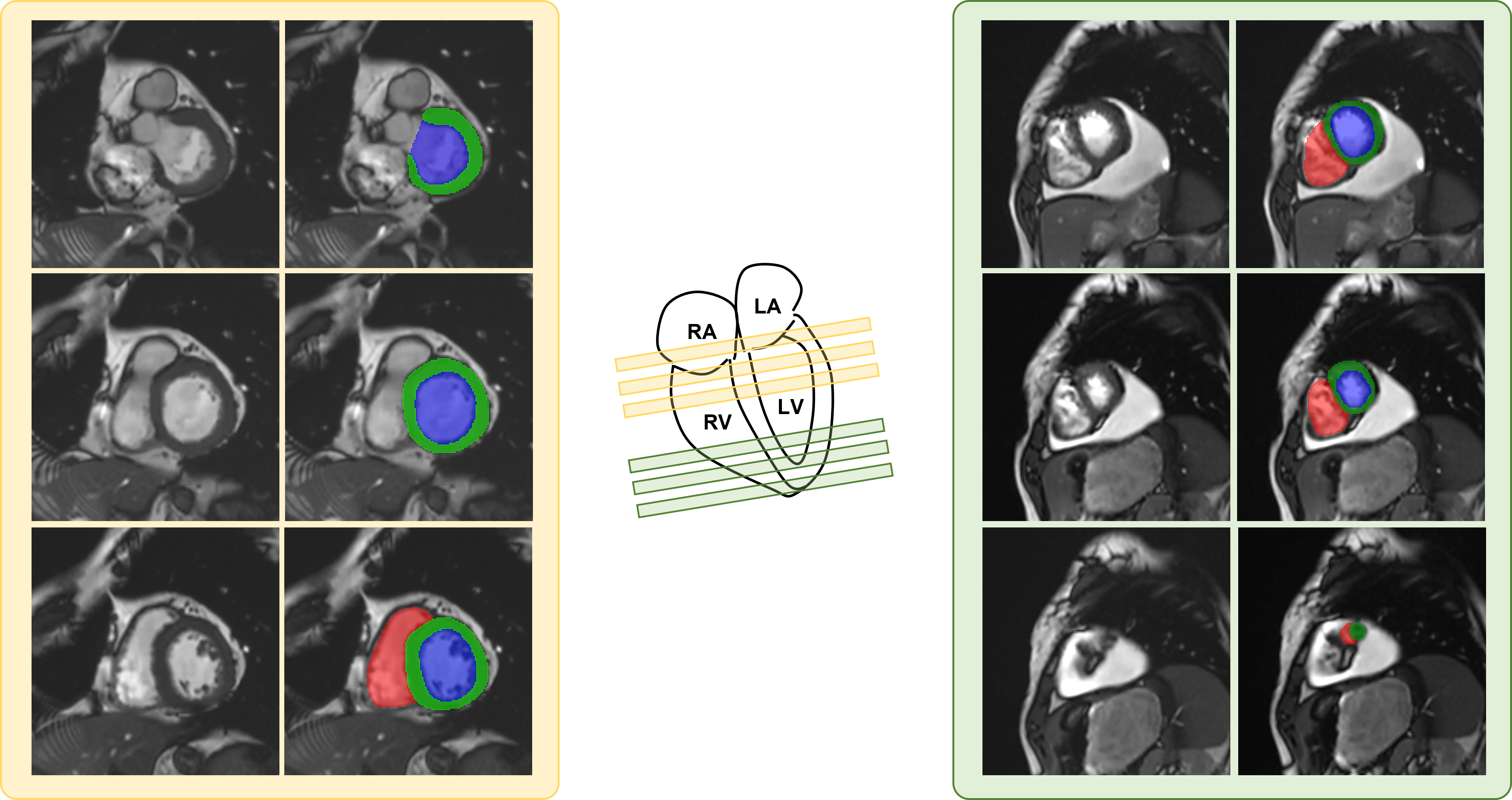}
\caption{Example of base (yellow) and apex (green) slices of CMR images with corresponding annotations. The red, green, and blue colors represent RV, MYO and LV, respectively. The middle diagram illustrates the locations of the slices. LV: left ventricle; RV: right ventricle; LA: left atrium; RA: right atrium.} \label{fig_challenge}
\end{figure}

Cardiac Magnetic Resonance imaging (CMR) is the gold standard for evaluating cardiac function by capturing heart structure and motion. Image segmentation is a crucial yet time-consuming step in CMR analysis, involving the segmentation of anatomies like the left ventricle (LV), right ventricle (RV), and myocardium (MYO). These segmentations provide clinical metrics such as volumes and volume ratios. However, CMR imaging presents challenges due to slow MRI acquisition and the need for breathholding~\cite{slomka2007patient}, resulting in multiple 2D image acquisitions with large slice gaps ($\sim$10mm). This introduces spatial discontinuity across slices, especially at boundaries like basal and apical slices (Fig.~\ref{fig_challenge}). Even for experienced radiologists, distinguishing ambiguous anatomies in these slices can be challenging due to the complex appearance variations.

Deep learning segmentation methods, such as \cite{sun2021saun,isensee2017automatic,isensee2021nnu,zotti2018convolutional,li2021cardiac,qin2018joint}, have demonstrated remarkable ability in automating the medical imaging segmentation process, including CMR. Convolution neural networks (CNNs), particularly U-Net like networks~\cite{ronneberger2015u,isensee2021nnu,wang2021sk} are a popular backbone in many applications. While 3D networks have the potential to combine information from different slices, previous work using 3D CNNs did not outperform 2D networks due to the challenge of modeling the large discontinuity across slices in CMR data \cite{patravali20172d,baumgartner2018ACDC}. Some approaches adopts a two-stage coarse-to-fine framework~\cite{ding2021tostagan,liu2020two} trying to fine-tune the coarse segmentation results. Several recent works in \cite{chen2021transunet,gao2021utnet,yan2022after,hatamizadeh2022swin,liu2022transfusion,cao2023swin} have combined the long-distance modeling capability of transformers with CNNs, demonstrating their superiority over pure CNN-based networks. Despite the advancements in CMR segmentation, segmenting the basal and apical slices remains challenging and impedes its broad clinical translation~\cite{chen2020deep}. In Fig.~\ref{fig_challenge}, we present a multi-slice CMR image that illustrates the variability in anatomy appearances across slices. The basal slices often include the chamber junction, where right atrium can appear very similar as RV. Determining apical segmentation also requires referring to neighboring slices due to the small region size.

Radiologists typically assess whether the anatomy is present in an image slice before proceeding with the annotation. Moreover, in challenging cases, radiologists may refer to the anatomy structures in the neighboring slices with clear information to aid in the annotation. Building upon these observations, we propose a framework that emulates the radiologists' annotation process, utilizing a two-stage and multi-task model structure with all-slice fusion transformer.

We conducted a comprehensive evaluation of our methods using both a public dataset (ACDC \cite{bernard2018deep}) and a private clinical CMR dataset. Our proposed method quantitatively outperformed both CNN-based and transformer-based models in terms of Dice coefficients. Notably, our approach produced significantly better results in challenging regions such as basal RV, apical RV, and basal MYO. In addition, qualitative assessments showed that our method generated visually appealing segmentation shapes that closely resembled human annotations without holes or broken fragments.

\section{Methods}

\begin{figure}[t]
\centering
\includegraphics[width=0.9\textwidth]{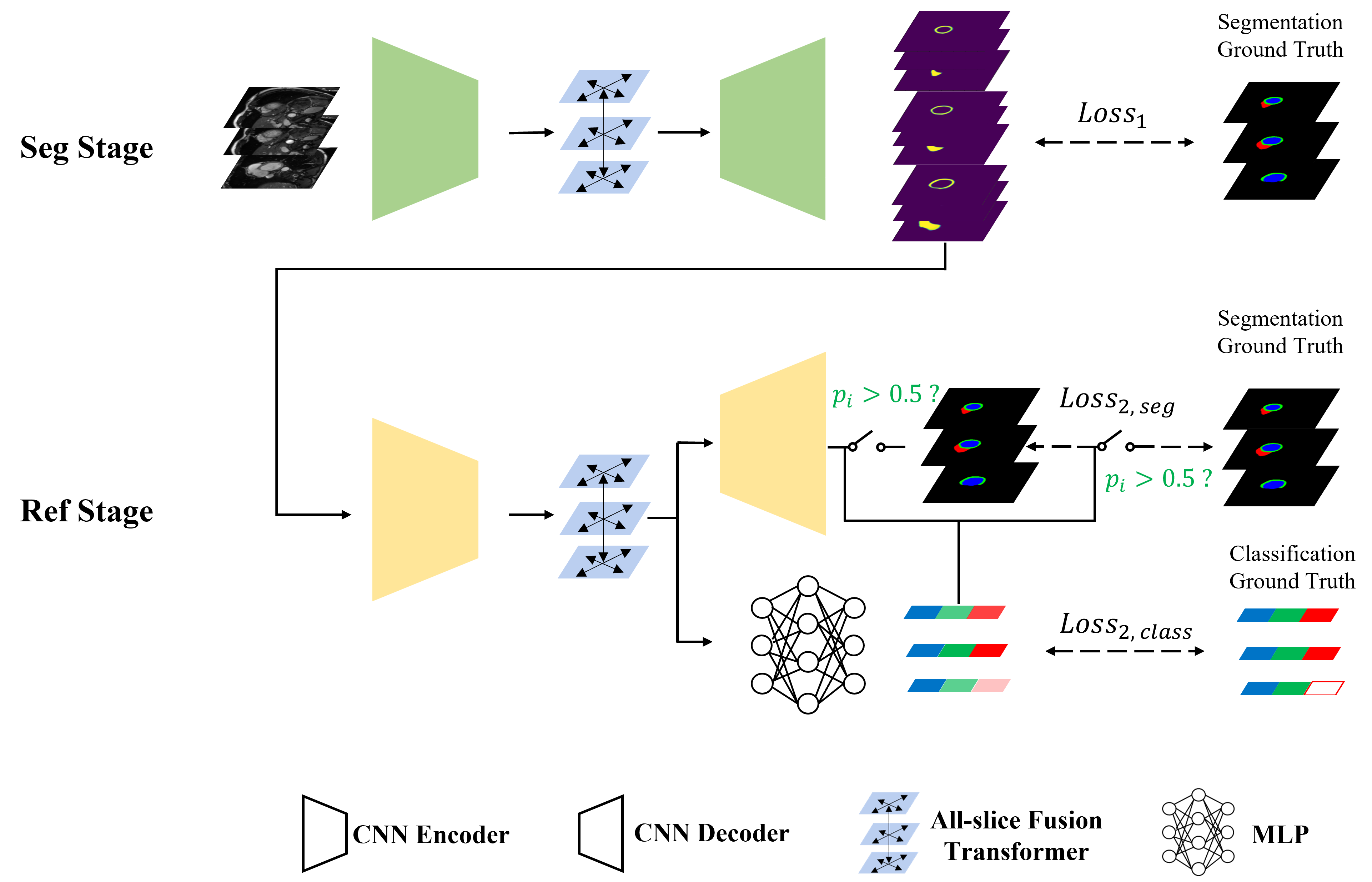}
\caption{Overview of the proposed two-stage and multi-task model with all-slice fusion transformer for cardiac imaging segmentation. The top and bottom parts are the initial segmentation stage (Seg stage) and segmentation refinement stage (Ref stage), respectively. The switches in Ref stage are turned on only when classification branch's output $p_{i}>0.5$.} \label{fig_all_structure}
\end{figure}

\begin{figure}[t]
\centering
\includegraphics[width=0.9\textwidth]{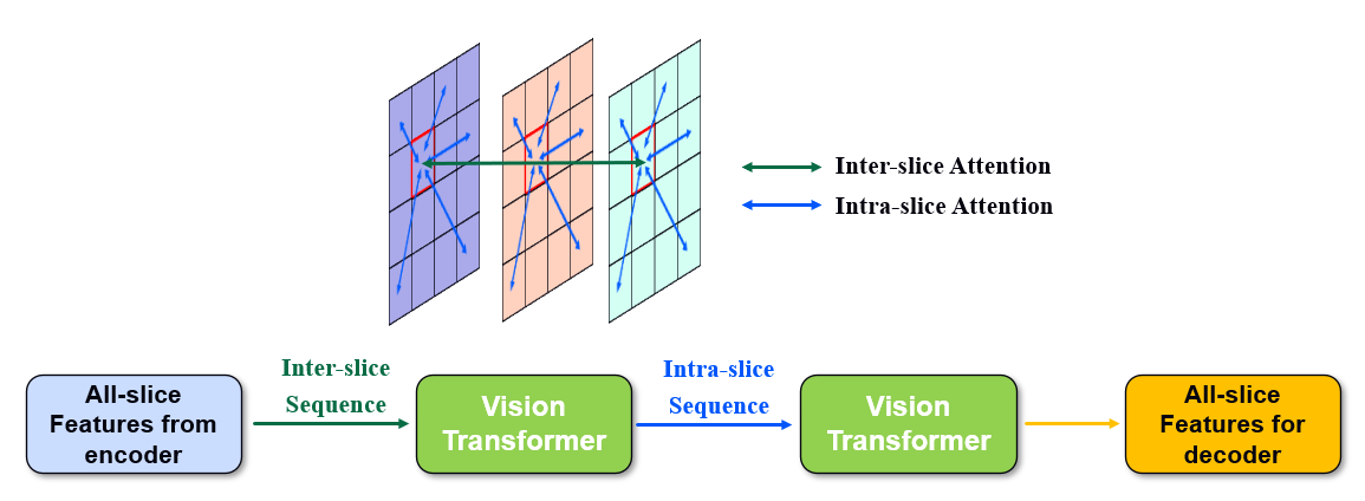}
\caption{The details of the all-slice fusion transformer bottleneck} \label{fig_transformer}
\end{figure}

\subsection{Two-stage based segmentation}

The proposed framework comprises two stages (Fig.~\ref{fig_all_structure}): initial segmentation (Seg stage) and segmentation refinement (Ref stage). During the Seg stage, a CNN + transformer model (detailed in section~\ref{transformer}) processes all the slices and generates initial segmentation probability maps. In the Ref stage, the probability maps from the Seg stage are inputted to a second CNN + transformer model, with the guidance of a classifier (explained in section~\ref{classifier}) to further improve the segmentation. Notably, while the first stage uses image information from all slices, the second stage does not require original images, simplifying the task and preventing variations in image intensities from affecting the results.

\subsection{All-slice fusion transformer} \label{transformer}
To efficiently fuse different slices' information and long-distance intra-slice information, we propose an all-slice fusion transformer that can take any number of slices as input and performs intra- and inter-slice attentions (Fig.~\ref{fig_transformer}). The transformers are placed at the bottlenecks of both Seg stage and Ref stage in Fig.~\ref{fig_all_structure}. Each stage's network is a hybrid of UNet~\cite{ronneberger2015u}
 with selective kernel (SK) convolutions~\cite{li2019selective,wang2021sk} and vision transformer~\cite{dosovitskiy2020image}. Images from all the slices are first concatenated along the batch dimension and then fed into a CNN encoder. Two consecutive transformers receive the bottleneck features from CNN encoder, and perform inter-slice attention and intra-slice attention respectively (Fig.~\ref{fig_transformer}). Lastly a CNN decoder with skip connections from the encoder produces segmentation results. This structure is motivated by AfterUNet \cite{yan2022after} but we add additional SK to enhance in-plane global feature extraction and also remove the position encoding in transformer to facilitate the variable input slice length.

\subsection{Classifier-guided segmentation refinement} \label{classifier}
The Ref stage is a multitasking (MT) model, consisting of a classification branch and a segmentation refining branch. The classification branch determines the presence or absence of specific anatomies while the segmentation refining branch predicts the final refined segmentation. The classification branch is a multilayer perceptron (MLP) which takes the bottleneck features from the CNN encoder and outputs a vector of [$p_{RV}$, $p_{LV}$, $p_{MYO}$], indicating the probability of the existence of RV, LV or MYO in the slice. The corresponding classification labels are derived from the ground truth segmentation. Note that the proposed classification branch is utilized to guide the segmentation refinement: 

\hspace{2em} 1) During training, only if the classification branch determines the presence of one class of anatomy ($p_{i}>0.5$), the segmentation output of that class is compared to the segmenatation ground truth in loss function. 

\hspace{2em} 2) During inference, only if the classification branch determines the presence of one class of anatomy ($p_{i}>0.5$), that class will be shown in the final segmentation output.

Under this training and inference scheme, the segmentation branch will only optimize and show high-certainty anatomies, reducing the chances of producing holes or fragments in segmentation. 
\subsection{Loss functions}
In Seg stage of Fig.~\ref{fig_all_structure}, the loss function $Loss_{1}$ is the cross-entropy loss between initial segmentation probability map and segmentation ground truth. In Ref stage, the loss function is a combination of both branches' loss: 

\[Loss_{2}=Loss_{2,seg} +\lambda~Loss_{2,class}=\sum_{i=1}^{N}\frac{1}{2}[sign(p_i-\frac{1}{2})+1]\cdot CE_{seg,i}+\lambda\sum_{i=1}^{N}CE_{class,i}\]


where $CE_{seg}$ is cross entropy loss between segmentation output and segmentation ground truth, $p_{i}$ is classification probability for a specific class, and $CE_{class}$ is cross entropy loss between classification probability and classification ground truth. $N=3$ is the number of total segmentation classes (RV, LV and MYO) and $\lambda=0.1$ was set to balance the loss from the two branches so that the scaling of two branches' losses were similar.

\section{Experiments}
\subsection{Datasets and settings}
The proposed method was evaluated using the ACDC dataset \cite{bernard2018deep} and a private CMR dataset collected from clinical routine with patient's consent. Both datasets contain multi-slice 2D cine CMR images where images were acquired at multiple slice locations and multiple time points. All slices at one time point is treated as a set for multi-slice experiments. For the ACDC dataset, each patient data has 6$\sim$18 slices and 2 time points. For the in-house dataset, each patient data has 4$\sim$18 slices and 6$\sim$29 time points. The training and testing data were split on the patient level. In total there are 8350/1859 images and 1123/198 multi-slice sets for training/testing. For the private data, each image was annotated LV, RV and MYO by two experienced annotators and was screened by an experienced doctor. The classification labels was calculated from the segmentation labels where an anatomy is present if the corresponding segmentation is present. For evaluation purpose, the stack of multiple slices were divided into base, mid and apex groups according to clinical standard, where in most cases the slices covering the whole heart were evenly divided into the three groups. 

All models were trained with an Adam optimizer and a learning rate of 0.0001 for 200 epochs. We trained all the models in PyTorch framework with a NVIDIA Tesla v100 GPU. As a part of the data preprocessing step, all images are resampled to a uniform in-plane spacing of 1.3 mm. Data augmentation techniques such as random rotation, shifting, flipping and cropping are employed. Dice score was used to evaluate segmentation results.

\subsection{Results}

\begin{table}[ht]
\caption{Comparison of methods in term of the Dice of RV, MYO and LV in different slice level. The bold texts mark the best performance.}\label{table1}
\small
\centering
\begin{tabular}{cccccc}
\hline
Models                                & Slice level & RV Dice        & MYO Dice       & LV Dice        & Average Dice   \\ \hline
\multirow{3}{*}{SS SK-UNet} & Base        & 0.909          & 0.847          & 0.939          & 0.898          \\
                                      & Mid         & 0.927          & 0.896          & 0.950          & 0.924          \\
                                      & Apex        & 0.800          & 0.807          & 0.917          & 0.842          \\ \hline
\multirow{3}{*}{AS After-UNet}  & Base        & 0.899          & 0.838          & 0.947          & 0.895          \\
                                      & Mid         & \textbf{0.932}       & \textbf{0.904} & \textbf{0.953}          & \textbf{0.930} \\
                                      & Apex        & 0.830          & 0.810          & 0.914          & \textbf{0.852}          \\ \hline
\multirow{3}{*}{\begin{tabular}[c]{@{}c@{}}Two-stage:\\ Det + Seg\end{tabular}}  & Base        & 0.907          & 0.843          & 0.940          & 0.897          \\
                                      & Mid         & 0.931          & 0.903          & \textbf{0.953}          & 0.929 \\
                                      & Apex        & 0.789          & \textbf{0.820}          & \textbf{0.922}          & 0.844          \\ \hline
\multirow{3}{*}{Proposed}             & Base        & \textbf{0.920} & \textbf{0.848} & \textbf{0.955} & \textbf{0.908} \\
                                      & Mid         & \textbf{0.932} & 0.902          & \textbf{0.953} & 0.929          \\
                                      & Apex        & \textbf{0.836} & 0.810 & 0.911          & \textbf{0.852} \\ \hline
\end{tabular}
\end{table}

The performance of the proposed method was compared with several state-of-the-art methods (Table.~\ref{table1}). For the single-slice (SS) network, SK-UNet \cite{wang2021sk} was used for comparison. For the all-slice (AS) network, After-UNet \cite{yan2022after} and a two-stage detection (Det) + segmentation (Seg) model \cite{liu2020two} were used for comparison. For all the models, the number of CNN blocks, as well as the corresponding convolution hyper-parameters are the same to make a fair comparison. 

As is shown in Table.~\ref{table1}, the proposed method outperforms all other methods on the average Dice of basal and apical slices. The proposed method's average Dice of mid slices (0.929) is almost the same to the best one (0.930). Regarding the difficult RV segmentation, the proposed method achieves the best RV Dice in all slice levels. The proposed method also achieves the best Dice scores for basal MYO, basal LV and mid LV. 
\begin{figure}[!ht]
\vspace{-0.5\baselineskip}
\centering
\includegraphics[width=0.98\textwidth]{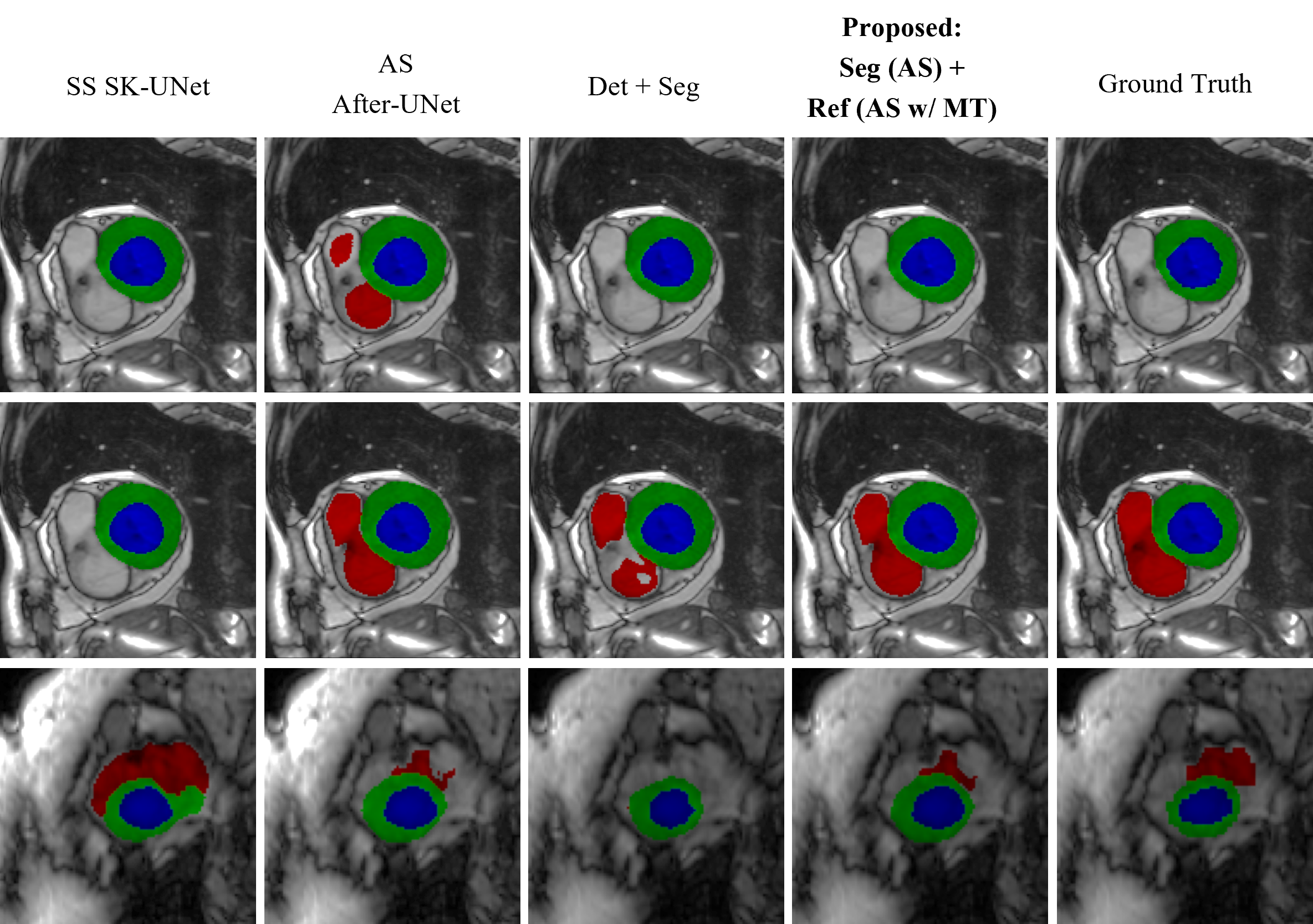}
\caption{Example segmentation results from state-of-the art methods, the proposed and the ground truth, at base (first and second rows) and apex (third row) slices.} \label{fig_imageresults}
\end{figure}
Fig.~\ref{fig_imageresults} shows example segmentation results for two basal slices from adjacent frames and one apical slice. As we can see in the ground truth column, although the two basal images look very similar, one has full RV annotations while the other has no RV annotation. The proposed method is the only method that can entirely match ground truth annotations and produce no holes or fragments. For the apical slice, the proposed method produces similar segmentation shape to the ground truth.

\subsection{Ablation study}

\begin{table}[ht]
\caption{The ablation study results. Seg: Initial segmentation stage; Ref: Segmentation refinement stage; MT: Multitasking model.}\label{table2}
\small
\centering
\begin{tabular}{ccccccc}
\hline
\multicolumn{2}{c}{Models}                                                                                                                                                                                             & Slice level & RV Dice & MYO Dice & LV Dice & Average Dice \\ \hline
\multicolumn{1}{c|}{\multirow{9}{*}{\begin{tabular}[c]{@{}c@{}}Single slice (SS)\\ models\end{tabular}}} & \multicolumn{1}{c|}{\multirow{3}{*}{Seg (SS)}}                                                              & Base        & 0.909   & 0.847    & 0.939   & 0.898        \\
\multicolumn{1}{c|}{}                                                                                    & \multicolumn{1}{c|}{}                                                                                       & Mid         & 0.927   & 0.896    & 0.950   & 0.924        \\
\multicolumn{1}{c|}{}                                                                                    & \multicolumn{1}{c|}{}                                                                                       & Apex        & 0.800   & 0.807    & 0.917   & 0.842        \\ \cline{2-7} 
\multicolumn{1}{c|}{}                                                                                    & \multicolumn{1}{c|}{\multirow{3}{*}{\begin{tabular}[c]{@{}c@{}}Seg (SS) + \\ Ref (SS w/o MT)\end{tabular}}} & Base        & 0.915   & 0.849    & 0.941   & 0.902        \\
\multicolumn{1}{c|}{}                                                                                    & \multicolumn{1}{c|}{}                                                                                       & Mid         & 0.929   & 0.898    & 0.949   & 0.925        \\
\multicolumn{1}{c|}{}                                                                                    & \multicolumn{1}{c|}{}                                                                                       & Apex        & 0.799   & 0.811    & 0.915   & 0.842        \\ \cline{2-7} 
\multicolumn{1}{c|}{}                                                                                    & \multicolumn{1}{c|}{\multirow{3}{*}{\begin{tabular}[c]{@{}c@{}}Seg (SS) + \\ Ref (SS w/ MT)\end{tabular}}}  & Base        & 0.914   & 0.856    & 0.952   & 0.907        \\
\multicolumn{1}{c|}{}                                                                                    & \multicolumn{1}{c|}{}                                                                                       & Mid         & 0.928   & 0.897    & 0.950   & 0.925        \\
\multicolumn{1}{c|}{}                                                                                    & \multicolumn{1}{c|}{}                                                                                       & Apex        & 0.795   & 0.810    & 0.916   & 0.840        \\ \hline
\multicolumn{1}{c|}{\multirow{12}{*}{\begin{tabular}[c]{@{}c@{}}All slice (AS)\\ models\end{tabular}}}   & \multicolumn{1}{c|}{\multirow{3}{*}{Seg (AS)}}                                                              & Base        & 0.899   & 0.838    & 0.947   & 0.895        \\
\multicolumn{1}{c|}{}                                                                                    & \multicolumn{1}{c|}{}                                                                                       & Mid         & 0.932   & 0.904    & 0.953   & 0.930        \\
\multicolumn{1}{c|}{}                                                                                    & \multicolumn{1}{c|}{}                                                                                       & Apex        & 0.830   & 0.810    & 0.914   & 0.852        \\ \cline{2-7} 
\multicolumn{1}{c|}{}                                                                                    & \multicolumn{1}{c|}{\multirow{3}{*}{Seg (AS + MT)}}                                                         & Base        & 0.893   & 0.839    & 0.945   & 0.892        \\
\multicolumn{1}{c|}{}                                                                                    & \multicolumn{1}{c|}{}                                                                                       & Mid         & 0.931   & 0.905    & 0.953   & 0.930        \\
\multicolumn{1}{c|}{}                                                                                    & \multicolumn{1}{c|}{}                                                                                       & Apex        & 0.809   & 0.817    & 0.916   & 0.847        \\ \cline{2-7} 
\multicolumn{1}{c|}{}                                                                                    & \multicolumn{1}{c|}{\multirow{3}{*}{\begin{tabular}[c]{@{}c@{}}Seg (AS) + \\ Ref (AS w/o MT)\end{tabular}}} & Base        & 0.914   & 0.848    & 0.952   & 0.905        \\
\multicolumn{1}{c|}{}                                                                                    & \multicolumn{1}{c|}{}                                                                                       & Mid         & 0.932   & 0.904    & 0.953   & 0.930        \\
\multicolumn{1}{c|}{}                                                                                    & \multicolumn{1}{c|}{}                                                                                       & Apex        & 0.834   & 0.813    & 0.911   & 0.853        \\ \cline{2-7} 
\multicolumn{1}{c|}{}                                                                                    & \multicolumn{1}{c|}{\multirow{3}{*}{\begin{tabular}[c]{@{}c@{}}Seg (AS) + \\ Ref (AS w/ MT)\end{tabular}}}  & Base        & 0.920   & 0.848    & 0.955   & 0.908        \\
\multicolumn{1}{c|}{}                                                                                    & \multicolumn{1}{c|}{}                                                                                       & Mid         & 0.932   & 0.902    & 0.953   & 0.929        \\
\multicolumn{1}{c|}{}                                                                                    & \multicolumn{1}{c|}{}                                                                                       & Apex        & 0.836   & 0.810    & 0.911   & 0.852        \\ \hline
\end{tabular}
\end{table}

Multiple ablation studies were performed to examine the effectiveness of our specific designs (Table.~\ref{table2}). We use SK-UNet as the backbone for SS models and After-UNet as the backbone for AS models, so Seg (SS) is SS SK-UNet, and Seg (AS) is AS After-UNet in Table.~\ref{table1}. By incorporating multi-slice information, the AS models show better segmentation performances in mid and apical slices.  By adding Ref stage, big improvements can be seen on challenging cases, such as RV, MYO and LV on the base slices and RV on apex slices. 

The additional classification branch (Seg(AS)+Ref(AS w/ MT)) further improves the very challenging cases such as the RV on the base level, agreeing with the qualitative example (Fig.~\ref{fig_imageresults}).  
In addition, by moving the classification branch to the Seg stage which takes the image level features as input (Seg (AS+MT)), we observed degraded quality comparing to Seg (AS). It suggests that the complex image features may deteriorate the classification accuracy and then damage the segmentation results. 

\section{Discussion and conclusion}

We proposed a two-stage, multi-task, and all-slice fusion transformer to address challenges in deep learning CMR segmentation of basal and apical slices. The two-stage and multi-task structures significantly enhance basal segmentation, while the all-slice fusion structure greatly improves apical and mid segmentation. Our method effectively leverages multi-slice information and the long-distance modeling capabilities of vision transformers, yielding promising results. The Ref stage, coupled with a classification branch, demonstrates notable improvements when incorporating multi-slice information, aligning with expert knowledge of referring to neighboring slices for delineating ambiguous anatomies. Additionally, our classifier-guided segmentation approach proves more effective when applied to coarse segmentation rather than the image itself. By removing intensity variations, artifacts, and adjacent anatomies in the image domain, the classifier-guided segmentation may focus more on the heart's structure and shape.

The classification branch serves as a crucial prior during both training and inference, providing global information to alleviate the limitations of pixel-level losses like cross-entropy and mean square error. The global supervision helps guide the network in handling ambiguous cases, such as the RV on base slices with even inconsistent ground truth labels. In this study, we focused on segmenting multi-slice cine images without leveraging temporal correlations, which could offer further advantages. However, the proposed features can be easily extended to explore temporal and spatiotemporal correlations. Our method is applicable to other multi-slice CMR applications, including Late Gadolinium Enhancement (LGE) images. Additionally, the choice of backbone is not restricted to After-UNet, and other backbones capable of effectively integrating 3D spatial information are worth investigating.


\textbf{Prospect of application}: The proposed study introduces a novel approach that aims to automate the segmentation of CMR images, thereby alleviating the workload burden on radiologists. Notably, the proposed work achieves accurate and visually pleasing segmentation shapes in challenging basal and apical slices, which may represent a significant stride towards the practical implementation of deep learning CMR segmentation in clinical settings.

%
%
%
\bibliographystyle{splncs04}
\bibliography{refs}

\end{document}